\documentclass[twocolumn,showpacs,aps,pra,preprintnumbers,amsmath,amssymb,superscriptaddress]{revtex4-1}

\usepackage{graphicx}
\usepackage{dcolumn}
\usepackage{bm}
\usepackage[usenames,dvipsnames]{color}
\usepackage{amsmath}
\usepackage[hypertex]{hyperref}
\usepackage{amsfonts}
\usepackage{amsthm}
\usepackage{amssymb}
\usepackage{mathrsfs}
\usepackage{subfigure}
\usepackage{ulem}

\begin{document}
\title{Conservation law
 for distributed entanglement of formation and quantum discord}

\author{Felipe F. Fanchini}
\email{fanchini@iceb.ufop.br}
\affiliation{Departamento de F\'{i}sica, Universidade Federal de Ouro Preto, CEP 35400-000, Ouro Preto, MG, Brazil}
\author{Marcio F. Cornelio}
\email{mfc@ifi.unicamp.br}
\affiliation{Instituto de F\'{\i}sica Gleb Wataghin, Universidade Estadual de Campinas, P.O. Box 6165, CEP 13083-970,
 Campinas, SP,
Brazil}
\author{Marcos C. de Oliveira}
\email{marcos@ifi.unicamp.br}
\affiliation{Instituto de F\'{\i}sica Gleb Wataghin, Universidade Estadual de Campinas, P.O. Box 6165, CEP 13083-970,
Campinas, SP,
Brazil}
\author{Amir O. Caldeira}
\email{caldeira@ifi.unicamp.br}
\affiliation{Instituto de F\'{\i}sica Gleb Wataghin, Universidade Estadual de Campinas, P.O. Box 6165, CEP 13083-970,
 Campinas, SP,
Brazil}

\date{\today}

\begin{abstract}
We present a direct relation, based upon a monogamic principle, between entanglement of formation (EOF) and quantum discord (QD), showing how they are distributed in an arbitrary tripartite pure system. By extending it to a paradigmatic situation of a bipartite system coupled to an environment, we demonstrate that the EOF and the QD obey a conservation relation. By means of this relation we show that in the deterministic quantum computer with one pure qubit the protocol has the ability to rearrange the EOF and the QD, which implies that quantum computation can be understood on a different basis as a coherent dynamics where quantum correlations are distributed between the qubits of the computer. Furthermore, for a tripartite mixed state we show that the balance between distributed EOF and QD results in a stronger version of the strong subadditivity of entropy.

\end{abstract}

 \maketitle

\section{Introduction}
Quantum discord (QD) is a measure of quantum correlation defined
by Ollivier and Zurek almost ten years ago \cite{ollivier} {and,  yet, a subject of increasing interest today} \cite{others}. It is
well known that, for a bipartite pure state, the definition of QD
coincides with that of the entanglement of formation (EOF). But it
has remained an open question how those two quantities would be
related for general mixed states. Here, we present this desired
relation for arbitrarily mixed states and show that the EOF and
the QD obey a monogamic relation. Surprisingly, this necessarily
requires an extension of the bipartite mixed system to its
tripartite purified version. Nonetheless, we obtain a conservation
relation for the distribution of EOF and QD in the system - the
sum of all possible bipartite entanglement shared with a
particular subsystem, as given by the EOF, cannot be increased
without increasing, \textit{by the same amount}, the sum of all QD
shared with this same subsystem. When extended to the case of a
tripartite mixed state, this relation results in a new
proof of the strong subadditivity of entropy, with stronger bounds
depending on the balance between the sum of EOF and the sum of QD
shared with a particular subsystem.

As an example of the importance of this conservation relation, we
explore the distribution of entanglement in the deterministic
quantum computation with one single pure qubit and a collection
of $N$ mixed states (DQC1). The algorithm, developed by Knill and
Laflamme \cite{knill1998},  is able to perform exponentially
faster computation of important tasks, \cite{al2,al3} when compared
with well-known classical algorithms, without any entanglement
between the pure qubit and the mixed ones \cite{al2}.  Arguably,
the power of the quantum computer is supposed to be related to QD,
rather than entanglement \cite{Datta08}. Here, using the
conservation relation, {we have shown} that even in the supposedly
entanglement-free quantum computation there is a certain amount of
multipartite entanglement between the qubits and the environment,
which is responsible for the non-zero QD (See Fig. 1).

\begin{figure}[h]
\begin{center}
\includegraphics[width=.48\textwidth]{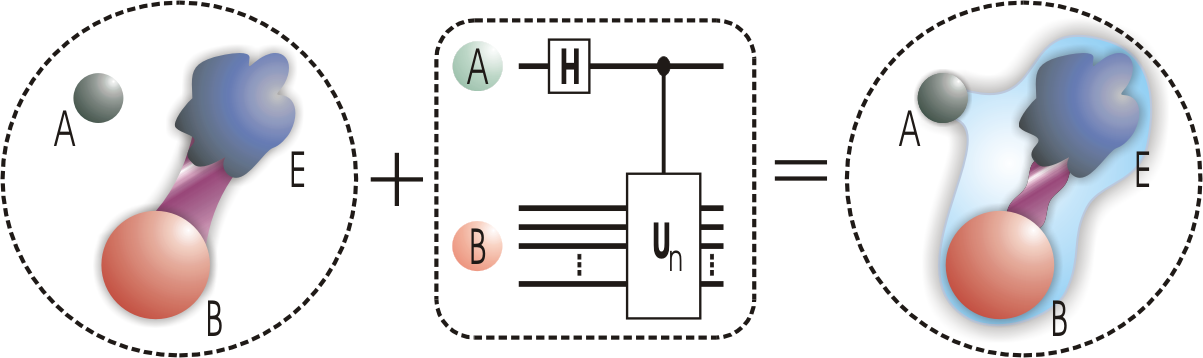} {}
\end{center}
\caption{(Color Online) Schematic illustration of the  DQC1. $A$ represents the pure qubit, $B$
the maximally mixed state, obtained through maximal entanglement
with the environment $E$. {From the left to the right, $B$ is initially entangled with $E$ (purple bar)}. The
protocol is then executed and $A$, although not directly entangled
with $B$, gets entangled with the pair $BE$ as the QD between $A$ and $B$ increase.}
\label {fig1}
\end{figure}

\section{Conservation relation}
Let us first consider an arbitrary system represented by a density
matrix $\rho_{ABE}$ with $A$ and $B$ representing two subsystems and $E$ representing the environment. It is important to emphasize that
the environment, here, is constituted by the universe minus the
subsystems $A$ and $B$, since, in this case, $\rho_{ABE}$ is a pure
density matrix. There is an important monogamic relation between
the entanglement of formation (EOF) \cite{bennett1996} and the
classical correlation (CC) \cite{henderson2001} between the two
subsystems developed by Koashi and Winter \cite{koashi2004}, {that}
we employ to understand the distribution of entanglement. It is
given by
\begin{equation}
E_{AB} + J^{\leftarrow}_{AE}=S_A, \label{koashi}
\end{equation}
where $E_{AB}\equiv E(\rho_{AB})$ is the EOF between $A$ and $B$,
$J^{\leftarrow}_{AE}\equiv J^{\leftarrow}(\rho_{AE})$ is the CC
between $A$ and $E$, and $S_A\equiv S(\rho_A)$ is the usual
Shannon entropy \cite{nielsen} of $A$. Further,
$\rho_{AB}=\rm{Tr}_E\left\{\rho_{ABE}\right\}$ and analogously for
$\rho_{AE}$ and $\rho_{A}$. Explicitly, CC reads $J^{\leftarrow}_{AE}=\max_{\{\Pi_x^E\}} \left[S(\rho_A) - \sum_x p_x S(\rho_{A}^x)\right]$
where the maximum is taken over all positive operator valued
measurements $\{\Pi^E_x\}$ performed on subsystem $E$, with
probability of $x$ as an outcome,
$p_x=\mbox{Tr}_A\left\{\Pi_x^{E}\rho_{AE}\Pi_x^{E}\right\}$  and
$\rho_{A}^x=\mbox{Tr}_E\{\Pi_x^E\rho_{AE}\Pi_x^E\}/p_x$. 
One can easily understand  Eq. (\ref{koashi}). The entropy $S(\rho_A)$
measures the amount of correlation (classical and/or quantum)
between $A$ {with} the external world. If we divide the external world into two parts, $B$ and $E$, the amount of quantum
correlation between $A$ and $B$, plus the amount of classical
correlation between  $A$ and the complementary part $E$, must be
equal to $S_A$. In this sense, Eq. (\ref{koashi}) poses
constraints on the ability that system $A$ has to share correlations
with other systems. For this reason it is called a monogamous
relation.

We can show a different aspect of Eq. (\ref{koashi}) by adding to
both of its sides the mutual information between $A$ and $E$,
$I_{AE}=S_A+S_E-S_{AE}$. After some manipulation we obtain
\begin{equation} 
E_{AB} = \delta^{\leftarrow}_{AE} + S_{A|E},\label{uni1}
\end{equation}
where $S_{A|E}=S_{AE}-S_E$ is the conditional entropy and
$\delta_{AE}^\leftarrow=I_{AE}-J_{AE}^\leftarrow$ is the QD
between subsystem $A$ and the environment $E$.  Eq. (\ref{uni1})
tells us that the entanglement between two arbitrary subsystems,
A and B, is related to the quantum discord between one of the
subsystems (A) and the environment E. It is important to note
that, although in Eq. (\ref{uni1}) the EOF is written as a
function of the QD between $A$ and $E$, it is straightforward to
write it as a function of the discord {between} $B$ and $E$. In
that case, $E_{AB} = \delta^{\leftarrow}_{BE} + S_{B|E}$. In the
same way, we can evaluate the QD between the subsystems $A$ and
$B$,
\begin{equation}
\delta^{\leftarrow}_{AB} = E_{AE} - S_{A|B},\label{uni2}
\end{equation}
which gives the quantum discord between A and B as a function of
the entanglement between A and E.  Remarking that, since the global state
is pure, $S_{A|B}=S_E-S_B$ which is in fact the EOF of the partition $E$ with $AB$, $E_{E(AB)}$,
minus the EOF of the partition $B$ with $AE$, $E_{B(AE)}$. Thus Eq. (\ref{uni2}) can be rewritten as
\begin{equation}
\delta^{\leftarrow}_{AB} = E_{AE} - E_{E(AB)}+ E_{B(AE)}.\label{uni21}
\end{equation}
This result shows that the EOF
and QD obey a very special monogamic relation, involving bipartite and tripartite entanglement.

We now derive a very simple but powerful result regarding
{the} distribution of bipartite entanglement.
Noting that $S_{A|B} = - S_{A|E}$ since $\rho_{ABE}$ is a pure
state and summing Eq. (\ref{uni1}) and Eq. (\ref{uni2}), we obtain
\begin{equation}
E_{AB} + E_{AE} = \delta^{\leftarrow}_{AB} + \delta^{\leftarrow}_{AE}.\label{unix}
\end{equation}

This important monogamic distribution of EOF and QD can also be viewed as a quantum
conservation law:

{\textit{Given an arbitrary tripartite pure system, the sum of all
possible bipartite entanglement shared with a particular
subsystem, as given by the EOF,  can not be increased without
increasing, by the same amount, the sum of all QD shared with this
same subsystem.}}

\section{Understanding the distribution of entanglement in the DQC1}
This last fundamental result has remarkable implications in the way
that entanglement can be distributed among many parties. For
example, we are now able to analyze the power of the quantum
computer ``without'' entanglement in view of this last statement.
In this sense, let us consider the DQC1 protocol, where the power of one
pure qubit was firstly revealed. It is well-known that any quantum
computation executed over $N$ maximally mixed states does not give
rise to exponential speedup when compared with the classical
computation. However, Knill and Laflamme \cite{knill1998}
demonstrated that if just one single pure qubit is added to this
set, the situation changes dramatically \cite{knill1998,al2,al3} -
For instance, the DQC1 gives an exponential speedup for the
computation of the normalized trace of an unitary operator,
$2^{-n} {\rm Tr}(U_n)$.

The DQC1 consists of a pure qubit that is represented here by the
subsystem $A$ and a completely mixed state of $n$ qubits,
$I_n/2^n$ that is represented by $B$. As illustrated in Fig.
(\ref{fig1}), we observe that initially the subsystem $A$ is pure
and has zero entanglement and zero discord with respect to $B$ and
$E$. On the other hand, the subsystem $B$ is given by the
maximally mixed state. It is important to emphasize here that even
a completely mixed state manifests its \textit{quantumness} by
the fact that it is impossible to distinguish the infinitely many
ensembles that can realize it. An alternative way to look at
this property is {to} consider it as an entangled state with an
external environment which {has} as many degrees of freedom as
necessary to purify the whole system. {Thus, we consider {here}
the {degree of mixture} of $B$ as due to {the} entanglement
between an environment $E$, which does not interact with $B$, and
$A$. {However}, it has interacted {with $B$} in the past, being
thus responsible for its \textit{mixedness}.
This approach 
 has been
fundamental for the understanding of important tasks in quantum information like Schumacher
compression, quantum state merging, and entanglement theory \cite{nielsen,Schumacher95}}.
Given this initial situation, we consider a circuit as that
exposed in Fig. (\ref{fig1}). We suppose that the subsystem $A$ is a qubit
in the initial state $|0\rangle$ (an eigenvector of the Pauli
matrix $\sigma_z$) and apply a Hadamard quantum gate, followed by
a control unitary on the remaining $n$ mixed qubit state. Thus,
after this process, the state of the subsystem $A$ and $B$ is
given by
\begin{eqnarray}
\rho_{AB}=\frac{1}{2^{n+1}}\left( \begin{array}{cc}
I_n & U_n^\dagger\\
U_n & I_n\end{array}\right),
\end{eqnarray}
since it gives a separated state with respect to $A$ and $B$
\cite{al2}. Expanding the state {on} the eigenstate basis
$\left\{|u_i\rangle\right\}$ of the unitary operator $U_n$ with
eigenvalues $\exp(i\theta_i)$ and considering the purifying
system eigenbasis $\{|e_i\rangle\}$, the joint $ABE$ state can be
written as
\begin{eqnarray}
|\psi_{ABE}\rangle=&\frac{1}{\sqrt{2^{n+1}}}&\sum_i(|0\rangle +
e^{i\theta_i}|1\rangle)\otimes |u_i,e_i\rangle.\label{rhodq}
\end{eqnarray}
Thus the expectation values of $\sigma_x$ and $\sigma_y$ on $A$
provide the normalized trace of $U_n$: $\left\langle
\sigma_x\right\rangle = {{\rm Re}\{ {\rm Tr}(U_n) \}}/{2^{n+1}}$
and $\left\langle \sigma_y\right\rangle = -{{\rm Im}\{ {\rm
Tr}(U_n) \}}/{2^{n+1}}$. At the end of the process, just before
the measurement that determines $\left\langle
\sigma_x\right\rangle$ and $\left\langle \sigma_y\right\rangle$,
we will have finite QD between $A$ and $B$, $
\delta^{\leftarrow}_{BA}$ \cite{Datta08}, but no entanglement
between them, $E_{AB}=0$\cite{al2}. Using the results given in Eq.
(\ref{uni2}) and Eq. (\ref{unix}), we meticulously examine the EOF
and the QD distribution. For this purpose, we examine how the
initial entanglement between $B$ and $E$ is affected during the
computation.
According to Eq. (\ref{uni2}), $ \delta^{\leftarrow}_{AB} = E_{AE}
- E_{E(AB)}+ E_{B(AE)}$, but from Eq. (\ref{rhodq}),  $E_{AE}=
\delta^{\leftarrow}_{AB}=0$. Similarly, we see that $E_{AB}=
\delta^{\leftarrow}_{AE}=0$, an so there really is no bipartite
entanglement between $A$ with $B$ or $E$. But in a similar fashion
to Eq. (\ref{uni21}) we can write
\begin{equation}
E_{BE} = E_{E(AB)} + \delta_{BA}^\leftarrow - E_{A(BE)}
\end{equation}
allowing us to analyze the EOF and the QD distribution in the
QDC1. Prior the computation,  $E_{BE}=E_{E(AB)}$, and
$\delta_{BA}^\leftarrow=E_{A(BE)}=0$. However, $\delta_{BA}^\leftarrow$ increases after the
controlled unitary operation,
implying necessarily in a redistribution of the entanglement
between the parties. In order {to} $\delta_{BA}^\leftarrow$ to
increase  some multipartite entanglement $E_{A(BE)}$ must exist.
This entanglement is indeed signaled by the mixed state of $A$
alone after the computation. Furthermore, using Eq. (\ref{unix}),
it is straightforward to show that for the DQC1 this entanglement
unbalance can be measured by the QD between $B$ and $E$, since
\begin{equation}
\delta_{BE}^\leftarrow = E_{E(AB)} - E_{A(BE)},
\end{equation}
and finally
\begin{equation}
\delta_{BA}^\leftarrow = E_{BE}-\delta_{BE}^\leftarrow.
\end{equation}
Nevertheless, it is important to emphasize that the power of the
quantum computer does not come only from the entanglement present
between $B$ and $E$, or even between $A$ and $BE$. In the DQC1, it
is clear that it comes from the protocol ability to redistribute
entanglement and quantum discord. This property is intrinsic of
the protocol and {does} not rely on the particularities of the
environment.  The DQC1 protocol ability to transfer entanglement
and its efficiency against classical algorithms for special tasks
can be tested in the light of the subsystem $B$ initial
entanglement with $E$. Had we started with a non-maximally mixed
state for $B$, meaning a non-maximally entanglement with $E$,
instead of Eq. (\ref{rhodq}), one would have ended up with
$|\psi_{ABE}\rangle=\sum_ic_i(|0\rangle +
e^{i\theta_i}|1\rangle)\otimes |u_i,e_i\rangle.$ In this case
$\langle\sigma_x\rangle$ or $\langle\sigma_y\rangle$ gives
$\textrm{Tr}[\rho_B U_n]$ \cite{knill1998}  containing, thus, less
information about the trace of $U_n$ when $B$ is initially less
entangled with $E$. The worst case is when $B$ is in a definite
state $|u_i\rangle$ (no entanglement with E), when we have access
to only one eigenvalue of $U_n$. Curiously this corresponds to the
situation where a maximal entanglement between $A$ and $B$ would
be available at the end, which certainly does not contribute to
any speedup for this special purpose. Therefore, we suggest that
one should look carefully at the redistribution of entanglement
during any quantum computation, and its implication for the
speedup of certain protocols. In the present situation, we see that
this ability for entanglement redistribution is a necessary (but
not sufficient) ingredient  for efficient quantum computation.

\section{Stronger bounds on the entropy strong subadditivity}

At this point, one could imagine what would be the implications of such a
relation  when some information is lacking for the description of
the global state, i. e., when the tripartite state involving
systems $A$, $B$, and $E$ is mixed.
In that case Eq. (\ref{koashi}) becomes an inequality
\cite{koashi2004} and, therefore,  Eq. (\ref{uni1}) turns into
\begin{equation} 
E_{AB} \le \delta^{\leftarrow}_{AE} + S_{A|E}. \label{u1}
\end{equation}
Similarly, by changing $B$ for $E$ in the equation above, it now
reads $E_{AE} \le \delta^{\leftarrow}_{AB} + S_{A|B}$, which when
added to Eq. (\ref{u1}) gives
\begin{equation}
S_B + S_E + \Delta \leq S_{AB} + S_{AE} \label{ineq}
\end{equation}
with
\begin{equation}
\Delta=E_{AB}+E_{AE}-\delta^{\leftarrow}_{AB} - \delta^{\leftarrow}_{AE}
\label{delta}
\end{equation}
being the balance between the entanglement and the quantum discord in the system.
The inequality (\ref{ineq}) can be stronger than the strong subadditivity (SS)  \cite{ruskai},
\begin{equation}
S_B + S_E  \leq S_{AB} + S_{AE},
\label{sub}
\end{equation}
depending on $\Delta$. For $\Delta>0$ it gives a
remarkable  lower bound for $S_{AB} + S_{AE}$,
{which is more restrictive} than (\ref{sub})
and must be fulfilled by any quantum system. Thus, we can define a more restrictive inequality than the
SS,
\begin{eqnarray}
S_B + S_E + \tilde{\Delta} \leq S_{AB} + S_{AE},
\label{newStrSub}
\end{eqnarray}
with $\tilde{\Delta}=\max\{0,\Delta\}$ where $\Delta$ is given by
the balance between EOF  and QD, Eq. (\ref{delta}).  
It is
important to emphasize that the SS, despite of being more difficult
to prove, is essentially derived through extensions of its
classical counterpart, but correlations play a different role in
quantum systems. So, it is not surprising that a more restrictive
bound may occur.

\begin{figure}
\begin{center}
\includegraphics[width=.48\textwidth]{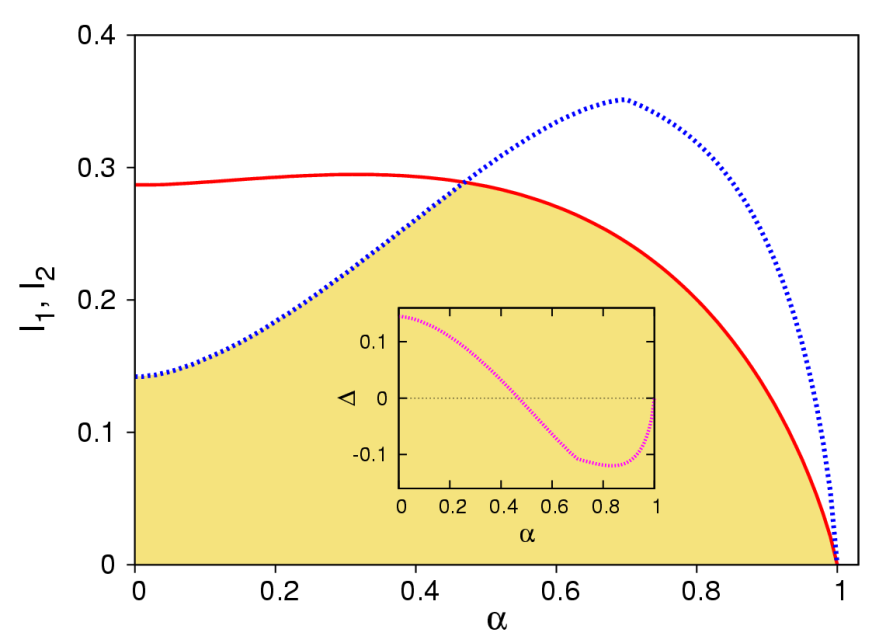} {}
\end{center}
\caption{(Color Online) The difference between the right and left
hand sides of Eq. (\ref{sub}), $I_1$, in red (solid line) and the
difference between the right and left hand sides of Eq.
(\ref{ineq}), $I_2$, in blue (dotted line). Combining these two
quantities the stronger inequality Eq. (\ref{newStrSub}) is
obtained. The difference between its right and left hand sides is
given by the shaded area.  The inset shows
$\Delta=E_{AB}+E_{AE}-\delta^{\leftarrow}_{AB} - \delta^{\leftarrow}_{AE}$.} \label {fig2}
\end{figure}

To exemplify this let us suppose we have a convex mixed state
$\rho_{ABE}=(1-\lambda)\frac{{\rm I}}{8}+\lambda\rho$,
where $\rho = |\Psi\rangle\langle \Psi|$:
$|\Psi\rangle = p\left[|101\rangle+|011\rangle\right] + \alpha|000\rangle$,
with $2p^2+\alpha^2=1$, and ${\rm I}$ is the identity operator
over the joint Hilbert space of $A+B+E$.  Let us define two
quantities
\begin{equation}
I_1=S_{AB} + S_{AE} - S_B - S_E \geq 0,
\end{equation}
and
\begin{equation}
I_2 = I_1 - \Delta \geq 0.
\end{equation}
In Fig. (\ref{fig2}) we plot $I_1$ and $I_2$ as a function of
$\alpha$, and, in the inset, we plot $\Delta$ for a  fixed
$\lambda=0.9$.  It is easy to see that in this situation $\Delta$
can be positive or negative. When $\Delta<0$ the  inequality given
by Eq. (\ref{ineq}) is weaker than the SS given by Eq.
(\ref{sub}).  However, when $\Delta>0$, meaning that the EOF of
all  bipartions is larger than their QD, Eq. (\ref{ineq}) is
stronger than Eq. (\ref{sub}), limiting the lower bound for
$S_{AB} + S_{AE}$. This is a {strikingly} different bound imposed
on the entropies of quantum systems, which is not shared by
{their} classical counterpart. The inequality above recovers the
SS only when $\tilde{\Delta}$ is null, meaning that the
distribution of bipartite entanglement is equal to the amount of
distributed quantum discord or smaller than that.
It is important to emphasize here the essential
role that SS plays in classical and quantum information theories
\cite{ruskai,newwinter}. Many fundamental inequalities, as
nonnegativity of entropy and subadditivity, can be derived from
that. To the best of our knowledge,  the only inequality known to
be independent is the one proposed in Ref. \cite{newwinter}, which
is valid when the SS saturates on some particular subsystems
configuration. It is straightforward to show that the inequality in Ref.
\cite{newwinter} is independent of (\ref{newStrSub}) as well, when
$\Delta>0$, since in this case SS can not be saturated.
However
something else can be learned from this saturation. Given a
quadripartite quantum system $\rho_{ABCD}$ such that SS is
saturated for the three triples $ABC$, $CAB$, and $ADB$ then,
$I_{CD} \geq I_{C(AB)}$  \cite{newwinter}. Substituting $I_{CD}$
by $\delta_{CD}^\leftarrow + J_{CD}^\leftarrow$ and using the
monogamic relation, Eq. (\ref{koashi}), and the conservation law,
Eq. (\ref{unix}), it is straightforward to show that when $I_{CD}
\geq I_{C(AB)}$ we have $E_{CD} - \delta^\leftarrow_{C(AB)} \geq
0$. So, as in Eq. (\ref{ineq}), the difference between the EOF and
the QD is of fundamental importance.

\section{Conclusion}
To summarize, we have given a monogamic relation between the EOF
and the QD. For that, we have derived a general interrelation on
how those quantities are distributed in a general tripartite system.
We applied {this} relation to show that in the DQC1 the entanglement present between one of the
subsystems and the environment is responsible for the non-zero quantum discord.
Since the maximally mixed
state is entangled with the environment, we show that the circuit
described by the DQC1 distributes this initial entanglement
between the pure qubit and the mixed state. 
Our results suggest that the protocol
ability to redistribute entanglement is a necessary condition for
the speedup of the quantum computer. 
In addition, we have extended
the discussion for an arbitrary tripartite mixed system showing
the existence of an inequality for the subsystems entropies which
is stronger than the usual SS.

We thank P. Hayden and K. Modi for helpful comments.
 This work is supported by FAPESP and CNPq through the
National Institute for Science and Technology of Quantum
Information.

\end{document}